\documentclass[twocolumn]{aastex701}

\usepackage{graphicx}	% Including figure files
\usepackage{amsmath}	% Advanced maths commands
\usepackage{float}
\usepackage{pgffor}   % For loops
\usepackage{ifthen}
\usepackage{comment}
\usepackage{subcaption}
\usepackage{float}
\usepackage{natbib}
\usepackage{tikz}
\usetikzlibrary{shapes.geometric, arrows.meta}
\usepackage{chngcntr}
\usepackage{rotating}

\begin{document}

%\title{An Extremely Low Surface Brightness Galaxy Candidate Discovered Near the Interacting Galaxy System NGC 4410 in Rubin LSST Early Data Preview 2}

\title{Rubin J122659.4+090236: An Extremely Low Surface Brightness Galaxy Candidate Discovered in the Rubin LSST Early Data Preview 2}

\author[0000-0002-8135-6249]{Dipanjan Mitra}
\affiliation{Inter-University Centre for Astronomy and Astrophysics, Ganeshkhind, Post Bag 4, Pune 411007, India}
\email[show]{dipanmitra1@gmail.com}  

\author[0000-0002-8768-9298]{Kanak Saha} 
\affiliation{Inter-University Centre for Astronomy and Astrophysics, Ganeshkhind, Post Bag 4, Pune 411007, India}
\email[show]{kanak@iucaa.in}

\author[0000-0002-5601-575X]{Sugata Kaviraj} 
\affiliation{Centre for Astrophysics Research, University of Hertfordshire, College Lane, Hatfield AL10 9AB, UK}
\email[]{s.kaviraj@herts.ac.uk}

%%%%%
\author[0000-0002-6496-9414]{Alister W. Graham}
\email{AGraham@swin.edu.au}
\affil{Centre for Astrophysics and Supercomputing, Swinburne University of Technology, Hawthorn, VIC 3122, Australia}
%%%%%

%%%%%
\author[0000-0002-8224-4505]{Sandro Tacchella}
\email{st578@cam.ac.uk}
\affil{The Kavli Institute for Cosmology (KICC), University of Cambridge, Madingley Road, Cambridge, CB3 0HA, UK}
\affil{Cavendish Laboratory, University of Cambridge, 19 JJ Thomson Avenue, Cambridge, CB3 0HE, UK}
%%%%%

%%%%%
\author[0000-0003-3921-2177]{Ricardo Demarco}
\email{ricardo.demarco@unab.cl}
\email{demarco.rj@gmail.com}
\affil{Institute of Astrophysics, Facultad de Ciencias Exactas, Universidad Andr\'es Bello, Sede Concepci\'on, Talcahuano, Chile}

%%%%%
\author[0000-0002-5678-1008]{Dieu D.\ Nguyen}
\email{dieun@umich.edu}
\email{nddieuphys@gmail.com}
\affil{Department of Astronomy, University of Michigan, 1085 South University Avenue, Ann Arbor, MI 48109, USA}
%%%%%

%%%%%
\author[0000-0002-5075-1764]{Elham Saremi}
\email{E.Saremi@soton.ac.uk}
\email{saremy.elham@gmail.com}
\affil{School of Physics and Astronomy, University of Southampton, Highfield Campus, Southampton SO17 1BJ, UK}
%%%%%

%%%%%
\author[0000-0001-9601-7779]{hector hernandez toledo}
\email{hector@astro.unam.mx}
\email{hector@astro.unam.mx}
\affil{Universidad Nacional Autónoma de México, Instituto de Astronomía, A.P. 70-264, 04510 CDMX, México}
%%%%%

%%%%%
\author[0000-0003-1896-0424]{Jose Benavides}
\email{jose.benavides@unc.edu.ar}
\email{jbenavid@ucr.edu}
\affil{Department of Physics and Astronomy, University of California, Riverside, 900 University Avenue, Riverside, CA 92521, USA}
%%%%%

%%%%%
\author[0000-0002-2238-9665]{Natanael M. Cardoso}
\email{natanael.mc@usp.br}
\email{nauxmac@gmail.com}
\affil{Escola Politécnica, Universidade de São Paulo, São Paulo, 05508-010, SP, Brazil}
%%%%%

\author[0000-0002-9091-2366]{Samuel Boissier}
\email{samuel.boissier@lam.fr}
\affil{Aix Marseille Univ, CNRS, CNES, LAM, Marseille, France}
%%%%%

%%%%%
\author[0000-0001-6268-1882]{Dan S. Taranu}
\email{dtaranu@astro.princeton.edu}
\affil{Department of Astrophysical Sciences, Princeton University, Princeton, NJ 08544, USA}

%%%%%
\author[0000-0002-4140-0110]{Bililign Dullo}
\email{dullob@erau.edu}  
\affil{Embry-Riddle Aeronautical University, Daytona Beach, FL 32114, USA}
%%%%%

%%%%%
\author[0000-0001-5242-2844]{Patricia B. Tissera}
\email{tissera.patriciab@gmail.com}
\affil{Instituto de Astrofísica, Pontificia Universidad Católica de Chile, Av. Vicuña Mackenna 4860, 7820436, Santiago Chile}
%%%%%

%%%%%
\author[0000-0003-2139-0944]{Matias Blaña}
\email{matias.blana.astronomy@gmail.com}
\affil{Centro Espacial Nacional, Fuerza Aérea de Chile, Av. Pedro Aguirre Cerda 5500, Cerrillos, Santiago, Chile}
%%%%%

%%%%%
\author[0009-0002-8051-1056]{Hao Fu}
\email{haofu@fudan.edu.cn}
\affil{Center for Astronomy and Astrophysics, Department of Physics, Fudan University, Shanghai 200438, China}
%%%%%

%%%%%
\author[0000-0002-4469-2518]{Priyanka Chakraborty}
\email{pchakraborty@uark.edu}
\affil{Department of Physics, University of Arkansas, 825 West Dickson Street, Fayetteville, AR 72701, USA
}
\affil{Center for Astrophysics, Harvard and Smithsonian, 60 Garden Street, Cambridge, MA 02138, USA}
%%%%%

%%%%%
\author[0000-0002-3963-3919]{Kevin A. Pimbblet}
\email{k.pimbblet@hull.ac.uk}
\affil{E.A.Milne Centre for Astrophysics, University of Hull, Cottingham Road, Kingston-upon-Hull, HU6 7RX, United Kingdom}
%%%%%

%%%%%
\author[0000-0003-3935-7018]{Tayyaba Zafar}
\email{tayyaba.zafar@mq.edu.au}
\affil{School of Mathematical and Physical Sciences, Macquarie University, NSW 2109, Australia}

%%%%%
\author[0009-0000-1088-4653]{Anirban Dutta}
\email{}
\affil{National Center for Nuclear Research, Pasteura 7, 02-093 Warsaw, Poland}

%%%%%
\author[0009-0004-1910-4990]{Aidan P Cotter}
\email{}
\affil{National Center for Nuclear Research, Pasteura 7, 02-093 Warsaw, Poland}

%%%%%
\author[0000-0001-5341-2162]{Darko Donevski}
\email{darko.donevski@ncbj.gov.pl}
\affil{National Center for Nuclear Research, Pasteura 7, 02-093 Warsaw, Poland}
\affil{INAF, Osservatorio Astronomico di Trieste, via Tiepolo 11, I-34131, Trieste, Italy}

%%%%%
\author[0000-0003-3080-9778]{Katarzyna Małek}
\email{Katarzyna.Malek@ncbj.gov.pl}
\affil{National Center for Nuclear Research, Pasteura 7, 02-093 Warsaw, Poland}
\affil{Aix Marseille Univ, CNRS, CNES, LAM, Marseille, France}

%%%%%
\author[0009-0000-6680-523X]{Rossella Ragusa}
\email{rossella.ragusa@inaf.it}
\affil{INAF – Osservatorio Astronomico di Capodimonte, Salita Moiariello
16, I-80131 Napoli, Italy}

%%%%%
\author[0000-0003-3080-9778]{Agnieszka Pollo}
\email{Agnieszka.Pollo@ncbj.gov.pl}
\affil{National Center for Nuclear Research, Pasteura 7, 02-093 Warsaw, Poland}
\affil{Astronomical Observatory of the Jagiellonian University, ul. Orla 171, 30-244 Kraków, Poland}

%%%%%
\author[0000-0002-2210-0681]{Unnikrishnan Sureshkumar}
\email{unni.suresh@ncbj.gov.pl}
\affil{National Center for Nuclear Research, Pasteura 7, 02-093 Warsaw, Poland}
\affil{Wits Centre for Astrophysics, School of Physics, University of the Witwatersrand, Private Bag 3, Johannesburg 2050, South Africa}

%%%%%
\author[orcid=0000-0002-2525-9647]{Aritra Ghosh}
\affiliation{Dept. of Astronomy \& the DiRAC Institute, University of Washington, Seattle, WA 98195, USA}
\email{aritraghsh09@gmail.com}

%%%%%
\author[0009-0005-3383-5551]{Antonio Vanzanella}
\email{Antonio.Vanzanella@ncbj.gov.pl}
\affil{National Center for Nuclear Research, Pasteura 7, 02-093 Warsaw, Poland}

%%%%%
\author[0000-0002-7413-0647]{Hareesh Thuruthipilly}
\email{hareesh.thuruthipilly@uva.es}
\affil{National Center for Nuclear Research, Pasteura 7, 02-093 Warsaw, Poland}
\affil{Departamento de Física Teórica, Atómica y Óptica, Universidad de Valladolid, 47011 Valladolid, Spain}
\affil{Laboratory for Disruptive Interdisciplinary Science (LaDIS), Universidad de Valladolid, 47011 Valladolid, Spain}

%\collaboration{3}{++Rubin Galaxy Science Collaboration}

%\collaboration{all}{The Terra Mater collaboration}


\begin{abstract}

We report the serendipitous discovery of an exceptionally low surface brightness galaxy (LSBG) candidate, Rubin J122659.4+090236, in Rubin Observatory imaging of the interacting NGC\,4410 system, identified in the \textit{Cosmic Treasure Chest} public release. 2D Sérsic modelling of the Rubin $g$, $r$, and $i$ images reveals a nearly round system with a shallow profile ($n\simeq0.4$), an effective radius of $\sim6''$, and central surface brightnesses of $\mu_{0,g}=27.52\pm0.04$, $\mu_{0,r}=27.62\pm0.07$, and $\mu_{0,i}=27.04\pm0.08$ mag arcsec$^{-2}$. EAZY photo-$z$ fitting favours an intermediate-$z$ solution at $z\simeq0.3$, while a low-redshift solution at $z\simeq0.028$, consistent with the NGC\,4410 system, is also permitted by a restricted EAZY fit over $0<z<0.1$ without imposing a redshift prior. These alternatives imply substantially different physical interpretations, ranging from a diffuse dwarf-like system to an exceptionally extended background LSBG. This discovery demonstrates Rubin's sensitivity to extremely diffuse galaxies and highlights the potential of the LSST survey to uncover large samples of such elusive systems across wide areas, enabling systematic studies of the LSBG population and its role in galaxy evolution.

\end{abstract}


\section{Introduction}
\label{sec:intro}

Low surface brightness galaxies (LSBGs) are among the most challenging galaxy populations to detect because much of their stellar light lies close to and well below the night sky background, with central surface brightnesses typically fainter than $\mu_{0,g}\gtrsim22.5$ mag arcsec$^{-2}$ in the $g$ band \citep{1997PASP..109..745B, 1999MNRAS.302L..55B}. Their diffuse nature has led to significant observational incompleteness in previous wide-field surveys \citep{1976Natur.263..573D, 1983MNRAS.205.1253D, 2025MNRAS.538..153K}, despite their importance for understanding galaxy formation in low-density regimes and the influence of environment on galaxy evolution \citep{1999MNRAS.302L..55B,2022MNRAS.513.3972Y,2024AA...682A...4T}. The most extended members of the LSBGs population are commonly referred to as ultra diffuse galaxies (UDGs). Recent work suggests that these systems are best viewed as the diffuse, large-size tail of the continuous LSBGs population rather than as a distinct galaxy class, with examples spanning environments from clusters to the field \citep{2001ApJ...556..177G, vanDokkum_2015, 2017MNRAS.468.4039R, 2018RNAAS...2...43C, 2024MNRAS.533.3771L, 2025PASA...42..155G}.

Over the past decade, increasingly deep imaging surveys have demonstrated that the low surface brightness population extends to substantially fainter surface brightnesses than those accessible to conventional wide-field galaxy catalogues. The Hyper Suprime-Cam Subaru Strategic Program \citep[HSC-SSP;][]{2018PASJ...70S...4A}, for example, identified 781 extended LSBGs in its first $\sim200$ deg$^2$ of imaging, with central $g$-band surface brightnesses extending to $\mu_{0,g}\simeq27.4$ mag arcsec$^{-2}$ \citep{Greco_2018}. The HSC-SSP sample includes a diverse population ranging from dwarf spheroidals and UDGs in nearby groups to gas-rich irregular galaxies and giant LSB spirals, demonstrating that the low surface brightness regime encompasses a broad range of galaxy properties. More recent deep imaging programmes have pushed to still fainter systems. The LBT Imaging of Galactic Halos and Tidal Structures (LIGHTS) survey, which targets the environments of nearby galaxies, reaches a surface-brightness depth of approximately $\mu_V\sim31$ mag arcsec$^{-2}$ and has revealed low-mass satellites with central surface brightnesses around $\mu_V(0)\sim27$ mag arcsec$^{-2}$ \citep{2021A&A...654A..40T}. Its more recent catalogue contains candidate satellite galaxies spanning approximately $24<\mu_{0,g}<28$ mag arcsec$^{-2}$, illustrating the ability of deep pointed observations to probe the extreme diffuse regime \citep{2024AJ....168...69Z}. These studies demonstrate that reaching very low surface brightness is possible with dedicated deep observations, but such surveys generally cover restricted regions of the sky or selected galaxy environments.

Recent discoveries further illustrate how far this parameter space extends. \cite{2026A&A...713L...8L} reported the serendipitous discovery of the low-mass galaxy TTT J1237327+143535 in the Virgo Cluster using the Two-meter Twin Telescope, with a central $g'$-band surface brightness of $\mu_{0,g'}=27.9\pm0.2$ mag arcsec$^{-2}$, an effective radius of $0.9\pm0.1$ kpc, and a stellar mass of only $(2.2\pm0.4)\times10^6 M_\odot$. This object occupies an extreme low surface brightness and low stellar mass regime and demonstrates that very diffuse dwarf like galaxies can be identified even with relatively modest telescope apertures when sufficiently deep imaging and careful analysis are employed. Similarly, \cite{2025ApJ...986L..18L} presented Candidate Dark Galaxy-2 in the Perseus Cluster, identified initially through an overdensity of globular clusters and subsequently supported by the detection of extremely faint diffuse emission in Hubble Space Telescope and Euclid imaging. Together, these studies demonstrate that the faintest diffuse systems can be uncovered through different observational strategies, including direct diffuse-light detection and indirect identification through their globular-cluster populations.

These observations highlight an important distinction between depth and survey area in constructing a census of the faintest galaxy populations. Deep pointed observations can reach extremely low surface brightnesses and reveal individual systems that would be difficult to identify in shallower imaging, but their limited or environmentally targeted footprints restrict the size and representativeness of the resulting samples. Conversely, a large-area survey provides the statistical volume required to identify rare objects and to investigate how their abundance and structural properties vary with environment. This distinction is particularly relevant for the lowest stellar mass and lowest surface brightness galaxies, for which small number statistics and selection effects can strongly influence the observed population. A sufficiently large-area sample can allow the distribution of size, luminosity, surface brightness, and colour to be studied together, while also providing comparisons between galaxies residing in the field, groups, and clusters. Such measurements are important for determining whether the extreme diffuse systems represent a continuous extension of the dwarf galaxy population and for understanding the physical processes that govern their formation and survival.

The Vera C. Rubin Observatory provides a complementary opportunity through the combination of deep multiband imaging and the very large area that will ultimately be surveyed by the Legacy Survey of Space and Time (LSST) \citep{brough2020verarubinobservatorylegacy, 2020arXiv200101728K}. Early Rubin observations have already demonstrated the recovery of extremely faint diffuse structures, including stellar streams \citep{Johnson_2026}, as well as the characterization of diffuse galaxies \citep{Romanowsky_2025}. The significance of Rubin for LSBG studies therefore need not rest on establishing a unique single-object surface-brightness limit. Rather, its major advantage is the ability to combine sensitivity to diffuse structures with a survey area large enough to identify much larger samples of rare systems. This will enable the extreme low surface brightness and low stellar-mass portions of the galaxy population to be explored statistically and across a much wider range of environments than is possible with individual deep pointed surveys. In this sense, the early Rubin detections provide a preview of the population studies that will become possible with the full LSST data set.

%The unprecedented depth and image quality of the Vera C. Rubin Observatory provide a major advance for studies of the low surface brightness Universe . Even the earliest public Rubin observations have demonstrated the observatory's sensitivity to the low surface brightness Universe, enabling the discovery of faint stellar streams  and the detailed characterization of diffuse galaxies \citep{Romanowsky_2025}. These initial studies highlight Rubin's potential to substantially expand the census of LSBGs while enabling detailed investigations of their stellar populations and environments.

%During a visual inspection of the Rubin Observatory \textit{Cosmic Treasure Chest} First Look field, centered on the nearby Virgo Cluster (M49) and released as part of Rubin Data Preview 2 (DP2), we identified a previously uncatalogued diffuse galaxy candidate projected near the interacting galaxy system NGC 4410. The object is either absent or only marginally visible in earlier optical surveys like SDSS \citep{York_2000}, DeCaLS \citep{2016AAS...22831701B}, emphasizing the sensitivity of Rubin imaging to extremely low surface brightness structures. Its proximity to an interacting group raises the possibility that it is either a previously unknown group member or a background diffuse galaxy.

In this work, we present the discovery and multiband characterization of a diffuse galaxy candidate Rubin J122659.4+090236 using Rubin Observatory imaging. We combine aperture photometry and spectral energy distribution modelling with 2D Sérsic fitting to constrain its photometric, stellar-population, and structural properties. We investigate its photometric redshift solutions and examine the implications of the redshift uncertainty for its intrinsic luminosity and physical size. Throughout this work, we adopt a flat $\Lambda$CDM cosmology with $H_0=67.4~{\rm km\,s^{-1}\,Mpc^{-1}}$ and $\Omega_{\rm m}=0.315$ \citep*{2020A&A...641A...6P}.

\begin{figure*}%[t!]
\centering
    \includegraphics[width=0.95\linewidth]{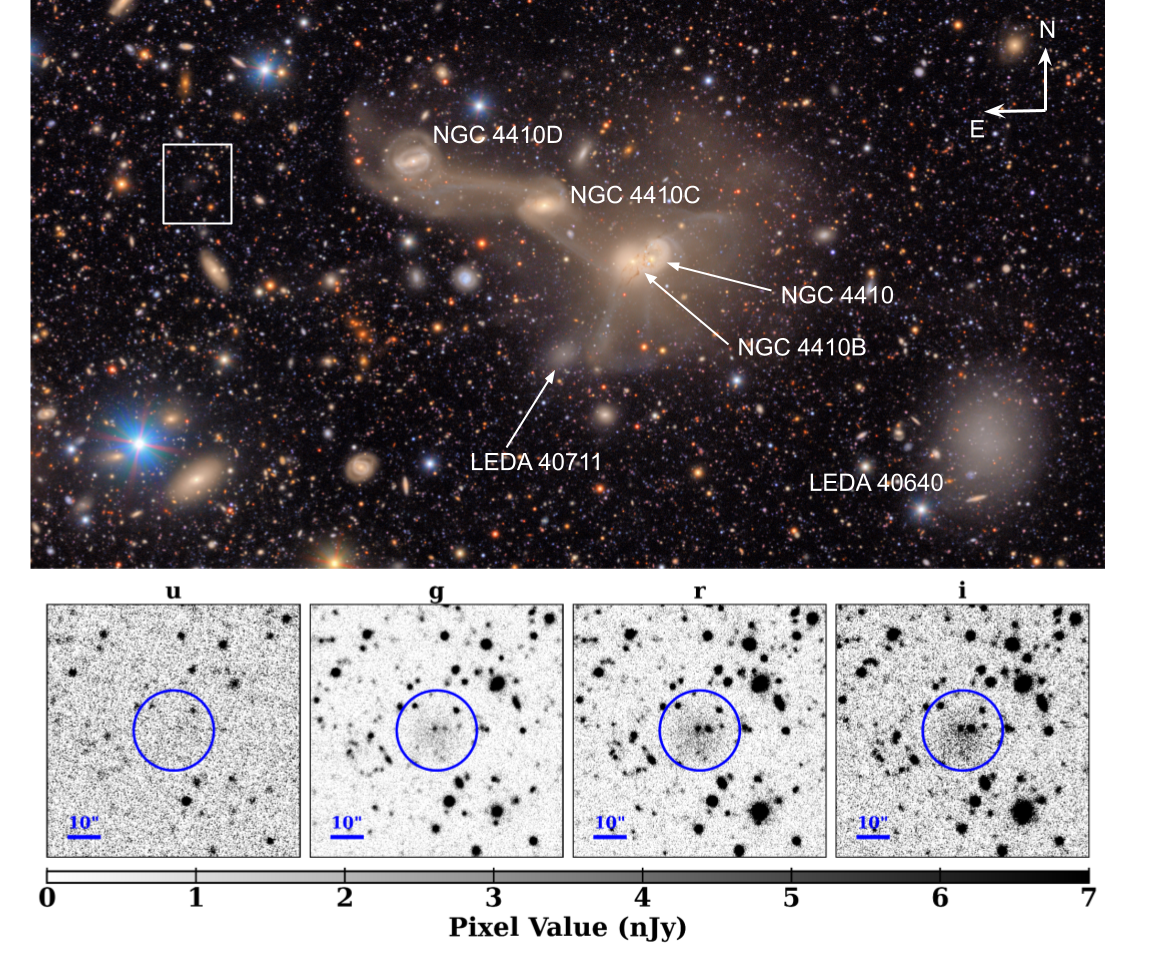}
    \caption{\textbf{Top:} Rubin Observatory colour-composite image of the interacting galaxy group NGC\,4410 in the Rubin First Look ``Cosmic Treasure Chest'' field. NGC\,4410/4410A, B, C, and D, together with the nearby low-surface-brightness galaxies LEDA\,40711 and LEDA\,40640 \citep{2023ApJS..267...27Z}, are labelled. The white box marks the location of the newly discovered LSB galaxy candidate presented in this work. The white compass denotes north and east. Image credit: NSF--DOE Vera C. Rubin Observatory/NOIRLab/SLAC/AURA. \textbf{Bottom:} $81''\times81''$ cutouts of the candidate in the $u$, $g$, $r$, and $i$ bands shown with a common linear intensity scale (nJy). The galaxy is clearly detected in the $g$, $r$, and $i$ bands but only marginally detected in the shallower $u$ band. The blue scale bar corresponds to $10''$.}
    \label{figfield}
\end{figure*}

\section{Data and METHODOLOGY}
\label{sec:data}
\subsection{Discovery and Rubin Data}

During a visual inspection of the Rubin Observatory \textit{Cosmic Treasure Chest}\footnote{\url{https://noirlab.edu/public/images/noirlab2521a/}} First Look field, centered on the nearby Virgo Cluster (M49) and released as part of Rubin Data Preview 2 (DP2), we identified a previously uncatalogued diffuse galaxy candidate, Rubin J122659.4+090236, projected near the interacting galaxy system NGC 4410. The visual inspection was exploratory rather than a completeness-calibrated search for LSBGs, and therefore the number of objects identified in this exercise cannot be used to estimate the completeness or limiting surface brightness of the Rubin data. In addition to genuine diffuse galaxies, visual searches at these surface-brightness levels can produce extended features associated with foreground stars, background galaxies or groups, Galactic cirrus, tidal debris, and local background fluctuations. We therefore treated the visual identification only as the initial candidate-selection step and subsequently subjected the candidate to multiband photometric, structural, and imaging checks.

Rubin J122659.4+090236 is located at RA = $186.74735^\circ$ and Dec = $9.04330^\circ$ (J2000), at a projected angular separation of $2.54'$ from NGC 4410D (Figure~\ref{figfield}, top panel). %It appears as a faint, smooth, and extended system, with no obvious central nucleus or prominent star-forming knots. 
It appears as a faint, smooth, and extended system with a very shallow light profile and no prominent star-forming knots. A compact, unresolved source is visible near the centre of the diffuse emission, but its physical association with the underlying galaxy cannot be established from the available imaging.
The source is not detected in earlier wide-field optical surveys, including SDSS \citep{York_2000} and DECaLS \citep{2016AAS...22831701B}, highlighting the sensitivity of Rubin imaging to extremely low surface brightness structures. Besides, the candidate is not detected in the H$\alpha$ imaging from the Virgo Environmental Survey Tracing Ionised Gas Emission \citep[VESTIGE;][]{2018A&A...614A..56B}. These non-detections do not by themselves establish the nature of the source, but provide useful constraints on possible interpretations. To assess whether the diffuse feature was associated with a local imaging artefact or a spurious background fluctuation, we independently verified the presence of diffuse emission at the same sky position using archival Canada-France-Hawaii Telescope (CFHT) MegaCam imaging \citep{2003SPIE.4841..513A} from the Next Generation Virgo Cluster Survey (NGVS; \citealt{2012ApJS..200....4F}) (see Appendix~\ref{app:cfht}). The proximity of the source to an interacting galaxy group raises the possibility that it is either a previously unknown group member or a background diffuse galaxy.

For our quantitative analysis, we use imaging from Rubin Data Preview 2 (DP2) \citep*{RTN-115}, obtained with LSSTCam \citep*{10.71929/rubin/2571927} at the Vera C. Rubin Observatory (Rubin:Simonyi) and processed using the Rubin Science Pipelines \citep*{10.71929/rubin/2570545}. The processed data products were retrieved from the Rubin U.S. Data Access Center (Rubin:USDAC) through the Rubin Science Platform. Calibrated coadded images in the $u$, $g$, $r$, and $i$ bands are provided in units of nJy pixel$^{-1}$ with a native pixel scale of $0.2''$ pixel$^{-1}$. The mean $3\sigma$ surface-brightness limits estimated within
$10\arcsec\times10\arcsec$ boxes in the coadded images are
$28.39$, $30.55$, $30.11$, and $29.45$~mag~arcsec$^{-2}$ in the
$u$, $g$, $r$, and $i$ bands, respectively.

\subsection{Image preparation and source masking}
\label{sec:imagemasking}

%We extracted $81\arcsec\times81\arcsec$ Rubin cutouts centered on Rubin J122659.4+090236, corresponding to approximately $405\times405$ pixels. We estimated and subtracted the local background using sigma-clipped statistics and identified contaminating sources with the Source Extraction and Photometry (SEP) package \citep{1996A&AS..117..393B,2016JOSS....1...58B}. The resulting source masks were used to exclude unrelated foreground and background sources while preserving the diffuse emission from the target. The image preparation and masking procedure is illustrated in Appendix~\ref{app:mask}.

%Two foreground stars projected on the source were retained rather than masked, since masking their pixels would also remove a substantial fraction of the underlying diffuse galaxy emission. Their contributions were instead modelled simultaneously with the galaxy during the two-dimensional structural fitting described in Section~\ref{sec:galfit}. The candidate is only marginally detected in the Rubin $u$ band (S/N $\simeq3.2$), whereas the $g$, $r$, and $i$ bands have substantially higher S/N. We therefore use all four bands for the Spectral Energy Distribution (SED) analysis, while restricting the structural modelling to $g$, $r$, and $i$.

\begin{figure*}
    \centering
    \includegraphics[width=\linewidth]{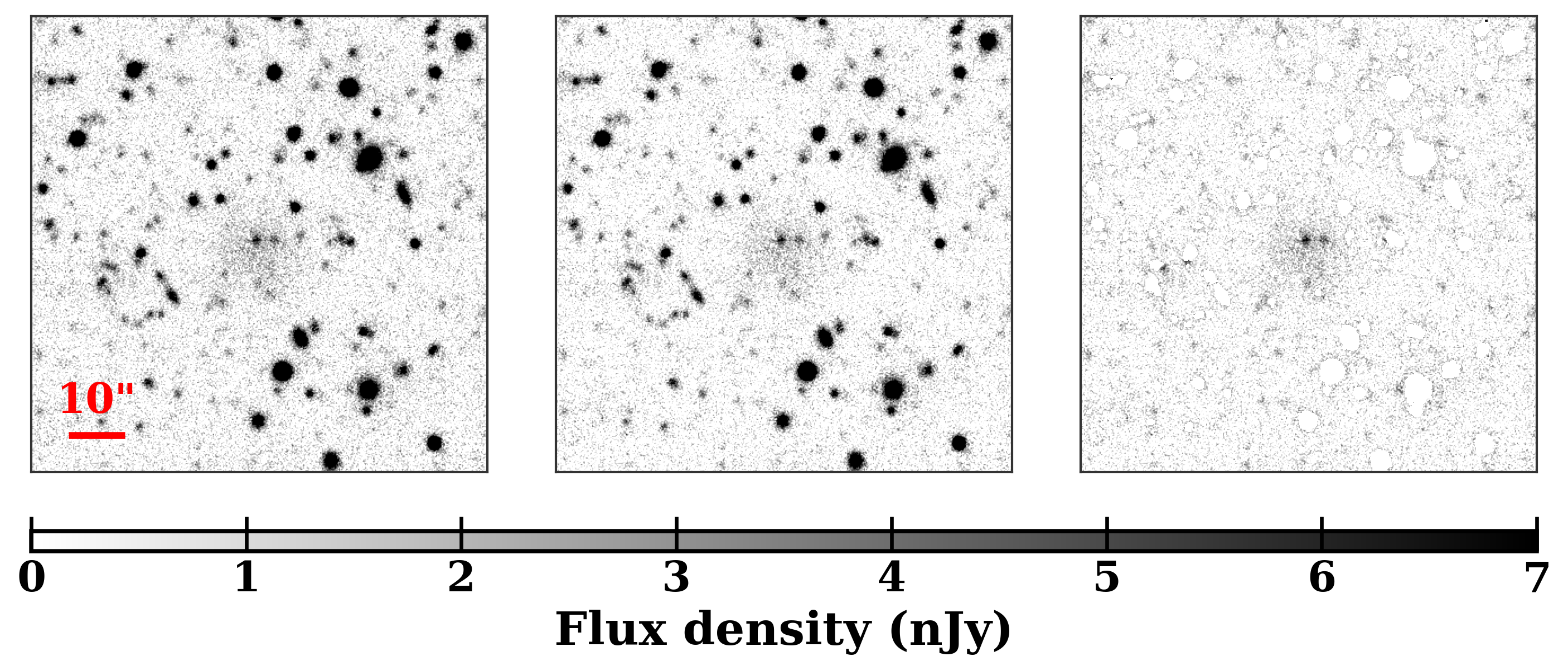}
    \caption{Image preparation and source masking for Rubin J122659.4+090236. From left to right, the panels show the original $g$-band $81\arcsec\times81\arcsec$ Rubin image, the background-subtracted image, and the final source-masked image used in the analysis. Contaminating foreground and background sources identified with SEP are masked, while the diffuse emission from the target is retained. The two foreground stars projected within $\simeq2.7\arcsec$ of the candidate are retained and modelled separately during the two-dimensional structural fitting. The $10\arcsec$ scale bar is shown in the left panel.}
    \label{fig:app_mask_1}
\end{figure*}

We extracted $81\arcsec\times81\arcsec$ Rubin cutouts centered on Rubin J122659.4+090236, corresponding to approximately $405\times405$ pixels. We first estimated and subtracted the local background using sigma-clipped statistics. Contaminating foreground and background sources were then identified using the Source Extraction and Photometry (SEP) package \citep{1996A&AS..117..393B,2016JOSS....1...58B}. The resulting segmentation map was used to mask unrelated sources while preserving the diffuse emission from Rubin J122659.4+090236. Pixels associated with the target were restored after source masking to ensure that the low-surface-brightness emission was not inadvertently removed. The image-preparation and masking procedure is illustrated in Figure~\ref{fig:app_mask_1}.
Two foreground stars projected directly onto the diffuse emission, at separations of approximately $2.7\arcsec$ from the candidate, were deliberately retained rather than masked. Masking these pixels would remove a substantial fraction of the underlying diffuse galaxy emission. Their contributions were instead modelled simultaneously with the galaxy during the two-dimensional \texttt{GALFIT} analysis described in Section~\ref{sec:galfit}. More distant contaminating foreground and background sources were masked. This treatment allows the diffuse emission to be retained while accounting explicitly for the two stellar sources that overlap the candidate.

As an independent assessment of the background properties and the effectiveness of the source masking, we also performed a random-aperture analysis on the original deep-coadd image. The details of this analysis and the resulting background-flux distributions are presented in Appendix~\ref{app:mask}.

The candidate is only marginally detected in the Rubin $u$ band (S/N $\simeq3.2$), whereas the $g$, $r$, and $i$ bands have substantially higher S/N. We therefore use all four bands for the spectral energy distribution (SED) analysis, while restricting the two-dimensional structural modelling to the $g$, $r$, and $i$ bands. Figure~\ref{fig:app_mask_1} shows the original, background-subtracted, and final source-masked $g$-band images used in the analysis.

\subsection{\texttt{GALFIT} Modelling}
\label{sec:galfit}

We modelled the 2D surface-brightness distribution of Rubin J122659.4+090236 with a single S\'ersic profile \citep{1963BAAA....6...41S} using
\texttt{GALFIT} \citep{2002AJ....124..266P}. The centroid, total magnitude, effective radius $R_{\rm e}$, S\'ersic index $n$, axis ratio $b/a$, and position angle were allowed to vary, while the sky background was fixed to the locally measured value described in Section~\ref{sec:imagemasking}. The source mask was used to exclude unrelated foreground and background
objects.

The two stars were modelled using PSF components. We used the position-dependent PSF models generated for the corresponding Rubin DP2 coadded images following the RSP tutorials provided by \cite{doecode_162015}. The PSFs were evaluated at the position of the candidate and supplied to \texttt{GALFIT}, with local FWHM values of $1.27\arcsec$, $1.16\arcsec$, and $1.21\arcsec$
in $g$, $r$, and $i$, respectively. Thus, each fit comprises one S\'ersic component for the diffuse galaxy and two PSF components for the nearby stars. This simultaneous modelling accounts for the stellar contamination without discarding pixels containing genuine low-surface-brightness emission. We assessed the fits from the best-fitting models and residual images, with particular attention to the low-surface-brightness outskirts
and the regions surrounding the two foreground stars.

%Because the diffuse emission of a low surface brightness galaxy can be comparable to the local sky background, we tested the sensitivity of our measurements to the adopted background level. First, as a check on the \texttt{GALFIT} modelling, we repeated the fit without subtracting the sky background beforehand and instead allowed \texttt{GALFIT} to fit the sky simultaneously with the galaxy and stellar components. The resulting structural parameters were then compared with those from our fiducial background-subtracted fit.

%We also assessed the effect of background uncertainty directly on the observed radial surface-brightness profile. In addition to the fiducial background-subtracted image, $I-S$, where $S$ is the locally estimated background, we constructed two images using $I-(S+0.5\sigma_{\rm bkg})$ and $I-(S-0.5\sigma_{\rm bkg})$, where $\sigma_{\rm bkg}$ is the RMS of the local background. The resulting radial profiles were used to illustrate the variation in the observed profile associated with a plausible change in the adopted sky level. This provides a direct visual assessment of the sensitivity of the measured low surface brightness emission to the local background uncertainty.

\subsection{SED Fitting}
\label{sec:sed}

We performed aperture photometry in the Rubin $u$, $g$, $r$, and $i$ bands using circular apertures of radius $2R_e$, with photometric uncertainties derived from the Rubin DP2 variance maps. We estimated the photometric redshift using \texttt{EAZY} \citep{2008ApJ...686.1503B} over $0<z<1$, both with and without the EAZY redshift prior. This broad range was adopted to avoid imposing a nearby interpretation based solely on the
candidate's projected proximity to NGC\,4410 and to test the sensitivity of the solution to the adopted prior. We additionally performed a restricted EAZY fit over $0<z<0.1$ to explicitly test the possibility that the candidate is associated with the NGC\,4410 system at $z\simeq0.02$ \citep{arp2006quasarsgalaxyclusterspaired}. The resulting redshift probability distributions are used to assess the plausible distance solutions; given the availability of only four optical bands, these should not be regarded as definitive redshift measurements. We subsequently fitted the observed $ugri$ SED with \texttt{CIGALE} \citep{2019A&A...622A.103B} to estimate the stellar population properties corresponding to the plausible redshift solutions. The CIGALE input parameters are listed in Appendix~\ref{apcig}. The resulting physical
properties are presented in Section~\ref{sec:results} and interpreted in Section~\ref{sec:discussion}.

\section{Results}
\label{sec:results}

\subsection{Structural Properties}
\label{sec:structural_properties}

\begin{figure*}%[t!]
\centering
    \includegraphics[width=\linewidth]{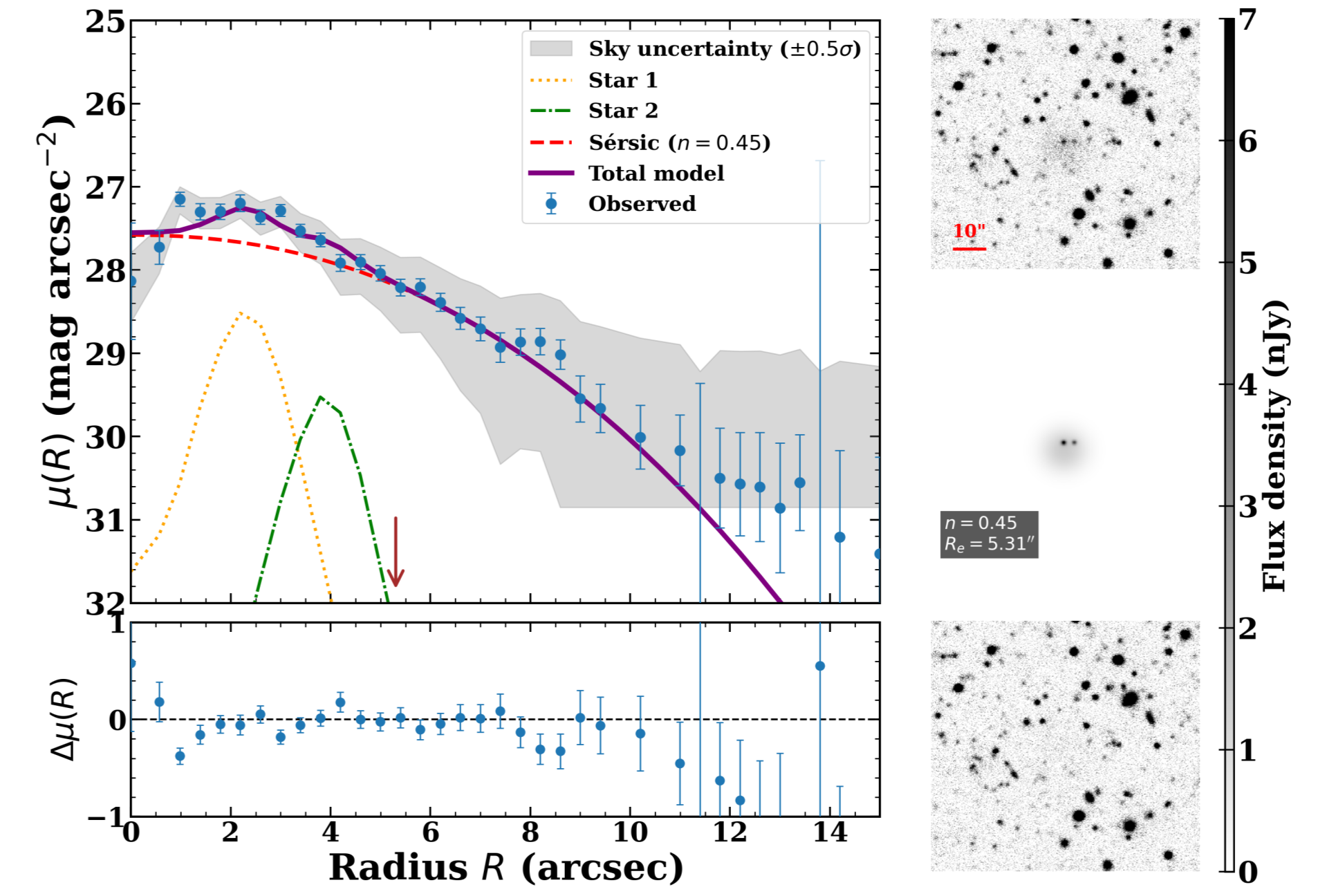}
    \caption{The $g$-band surface-brightness profile and two-dimensional
\texttt{GALFIT} model of Rubin J122659.4+090236. Blue points show the observed azimuthally averaged profile, with the shaded region indicating the change resulting from a $\pm0.5\sigma$ variation in the adopted sky level. The red dashed curve is the best-fitting S\'ersic component ($n=0.45$), while the orange dotted and green dash-dotted curves show the two simultaneously fitted foreground stars. The purple curve shows the total model. The lower panel gives the residual profile, and the vertical
arrow marks $R_{\rm e}=5.31\arcsec$. The right panels show the image, best-fitting model, and residual, respectively.}
    \label{figgalfit}
\end{figure*}

Figure~\ref{figgalfit} shows the 2D \texttt{GALFIT} modelling
of Rubin J122659.4+090236 in the $g$ band, together with the corresponding radial surface-brightness profile and residuals. The galaxy is well described by a shallow Sérsic component, while the two nearby foreground stars are simultaneously modelled using PSF components. The residuals remain small over most of the radial range in which the galaxy is detected,
indicating that the diffuse emission is well reproduced by the adopted model.

The resulting parameters for independent $g$, $r$, and $i$ image fitting are summarized in Table~\ref{tab:galfit_results}. The candidate is extended in all three
bands, with effective radii of $R_{\rm e}=5.31\pm0.18\arcsec$, $5.93\pm0.17\arcsec$, and $5.60\pm0.21\arcsec$ in $g$, $r$, and $i$, respectively. The Sérsic indices are consistently low, $n=0.45\pm0.03$, $0.43\pm0.02$, and $0.36\pm0.03$, indicating an unusually shallow light profile. The fitted axis ratios of $b/a=0.99\pm0.03$, $0.87\pm0.02$, and $0.84\pm0.03$ show
that the system is broadly round, with a modest increase in ellipticity toward longer wavelengths. The central surface brightness is extremely faint, with $\mu_{0,g}=27.56\pm0.09$,
$\mu_{0,r}=27.04\pm0.08$, and $\mu_{0,i}=26.78\pm0.11$~mag~arcsec$^{-2}$. The corresponding mean surface brightness within the effective radius is
$\langle\mu\rangle_{e,g}=27.80\pm0.08$,
$\langle\mu\rangle_{e,r}=27.25\pm0.07$, and
$\langle\mu\rangle_{e,i}=26.89\pm0.09$~mag~arcsec$^{-2}$. The combination of large angular extent, very low surface brightness, and $n<0.5$ in all three bands demonstrates the exceptionally diffuse nature of the candidate.

We assessed the sensitivity of the observed profile to the local sky uncertainty by recomputing the background-subtracted images using sky levels of $S\pm0.5\sigma_{\rm sky}$. These perturbed profiles were not refitted; instead, they provide an estimate of the variation in the observed surface brightness associated with the background uncertainty. As shown by the shaded region in Figure~\ref{figgalfit}, the resulting profiles remain consistent with the best-fitting Sérsic model over the radial range where the galaxy is detected. We therefore find that the
observed diffuse structure is robust against reasonable variations in the adopted sky level. As an additional check, we repeated the \texttt{GALFIT} modelling while allowing
the sky background to vary simultaneously with the galaxy model (for details, see Appendix~\ref{galskyap}). The resulting structural parameters are consistent with those obtained from the sky-subtracted images (see Table~\ref{tab:galfit_results}).

\subsection{Redshift and Physical Properties}
\label{sec:results_sed}

\begin{figure*}%[t!]
\centering
    \includegraphics[width=\linewidth]{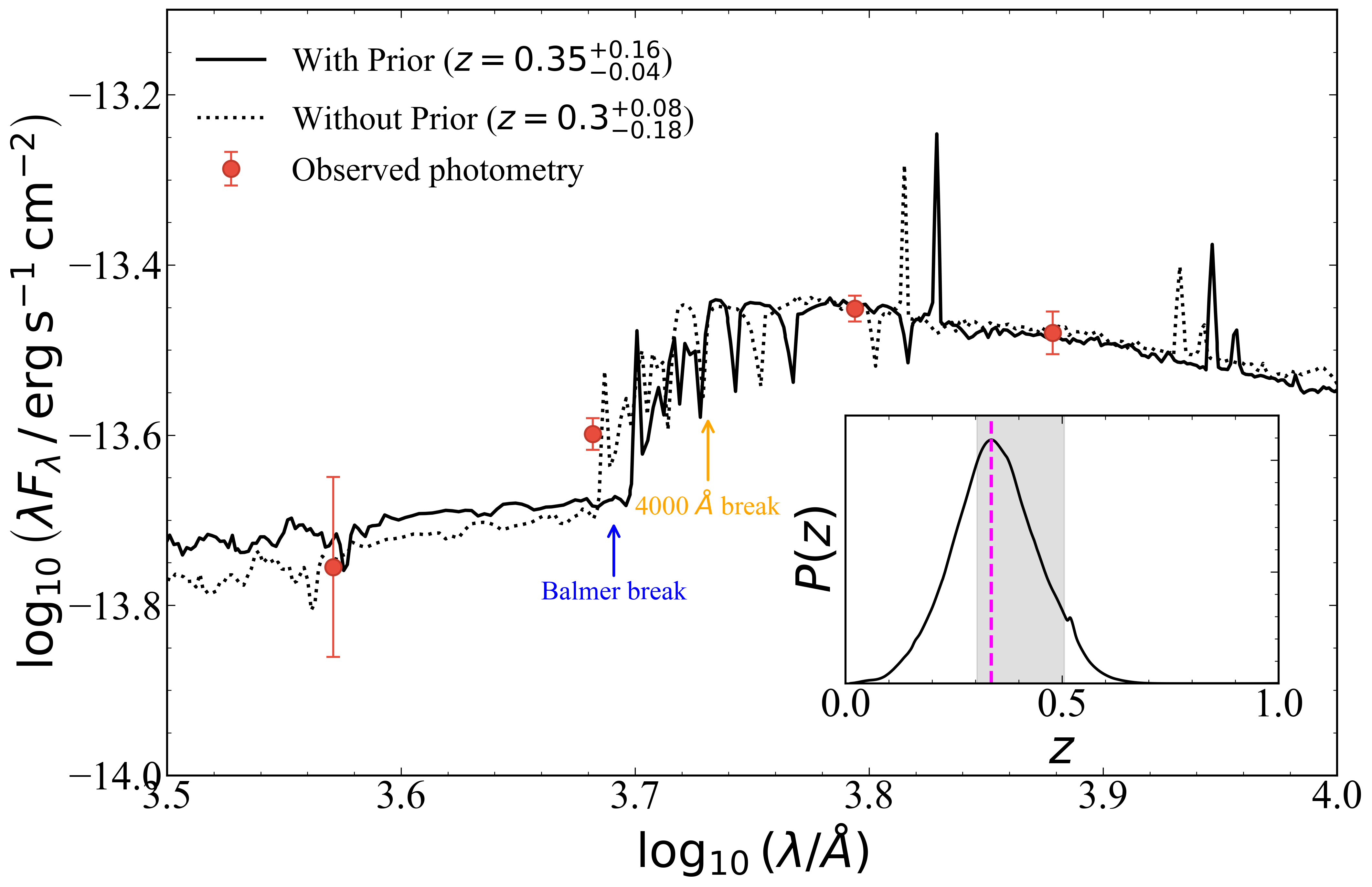}
    \caption{Best-fit EAZY SEDs for Rubin J122659.4+090236. The observed $u$, $g$, $r$, and $i$ photometry is shown by the red points with uncertainties. The solid and dotted black curves show the best-fitting EAZY templates obtained with and without the redshift prior, respectively, yielding $z=0.35^{+0.16}_{-0.04}$ and $z=0.30^{+0.08}_{-0.18}$. The locations of the Balmer and 4000\,\AA\ breaks at the reference redshift are indicated. The inset shows the corresponding EAZY redshift probability distribution $P(z)$, with the vertical dashed line marking the peak of the adopted unrestricted solution at $z=0.35$. The broad and asymmetric $P(z)$ illustrates the substantial redshift uncertainty resulting from the limited four-band photometry.}
    \label{figsed}
\end{figure*}

\begin{figure*}[t!]
    \centering
    \includegraphics[width=\linewidth]{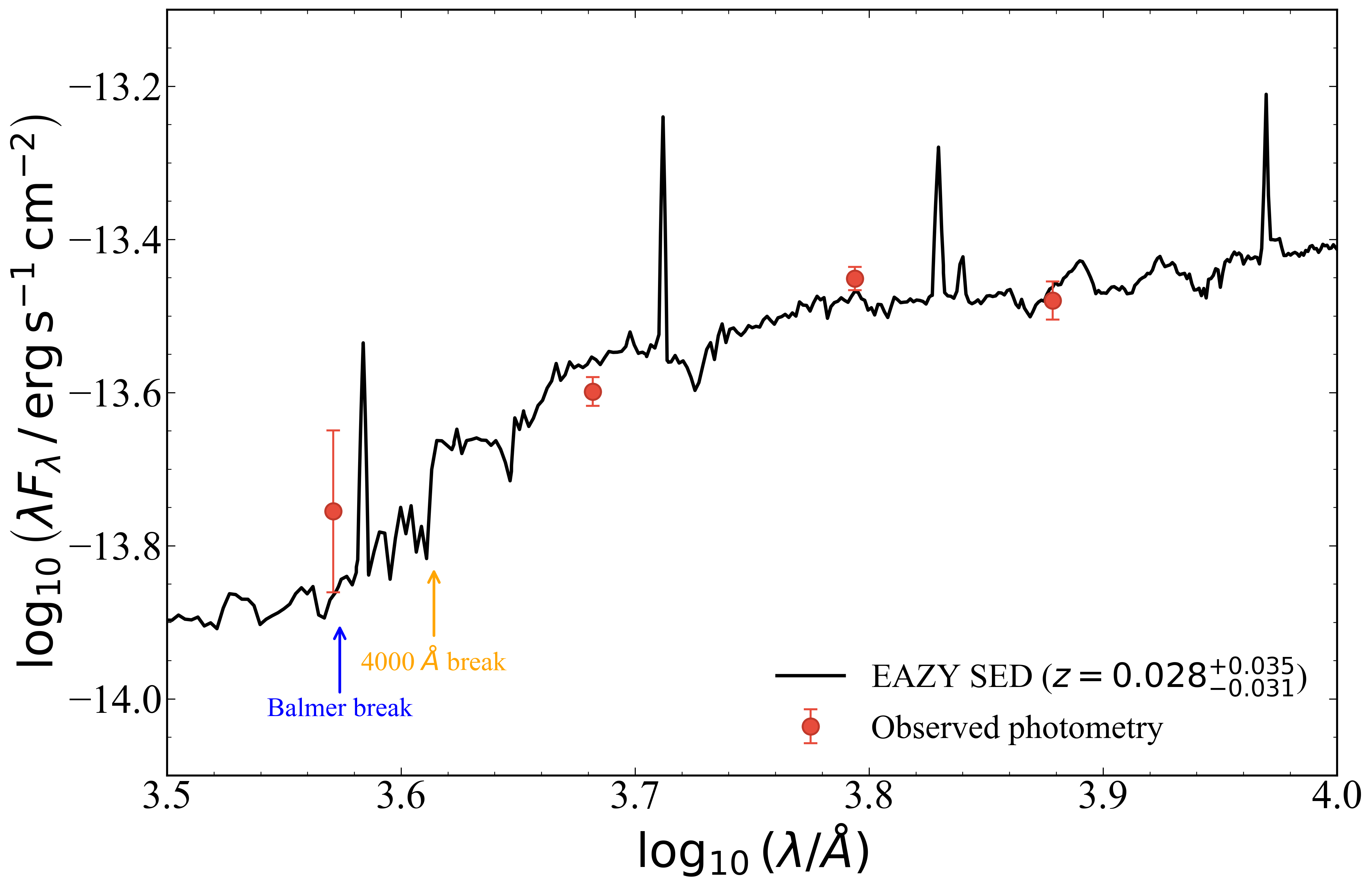}
    \caption{Best-fitting EAZY SED for Rubin J122659.4+090236 from the restricted $0<z<0.1$ fit without a redshift prior. The black curve shows the best-fitting EAZY template at
    $z=0.028^{+0.035}_{-0.031}$, while the red points show the observed Rubin $ugri$ photometry. The locations of the Balmer and 4000~\AA\
    breaks are indicated. The narrow features in the template correspond to nebular emission lines expected at this redshift, including
    [O\,II], H$\beta$/[O\,III], and H$\alpha$; these are template features rather than spectroscopic detections.}
    \label{fig:eazy_lowz}
\end{figure*}

The photo-$z$ and SED analysis permits three relevant solutions for Rubin J122659.4+090236. For the unrestricted EAZY run $0<z<1$, both without and with prior, we obtain $z=0.30^{+0.08}_{-0.18}$ and $z=0.35^{+0.16}_{-0.04}$, respectively. The resulting EAZY SED and $P(z)$ distributions are shown in Figure \ref{figsed}. For the restricted run $0<z<0.1$, we obtain $z_{\rm peak}\simeq0.028^{+0.035}_{-0.031}$. The best fit SED is shown in Figure \ref{fig:eazy_lowz}. All 3 solutions provide plausible descriptions of the available four-band photometry, and the broad, asymmetric probability distributions highlight
the limited redshift constraint. We therefore treat these solutions as distinct, viable scenarios rather than adopting a single definitive photometric redshift.

For the two intermediate-$z$ solutions, the corresponding CIGALE fits give broadly similar physical properties. At $z=0.30$, we obtain $M_\star=(2.98\pm1.28)\times10^{9}\,M_\odot$
[$\log_{10}(M_\star/M_\odot)=9.47\pm0.19$] and
${\rm SFR}=0.74\pm0.25\,M_\odot\,{\rm yr}^{-1}$, while at $z=0.35$, the inferred values are
$M_\star=(3.48\pm1.57)\times10^{9}\,M_\odot$ and
${\rm SFR}=1.14\pm0.41\,M_\odot\,{\rm yr}^{-1}$. Thus, either
intermediate-$z$ solution corresponds to a galaxy with a stellar mass of a few $10^9\,M_\odot$ and modest ongoing star formation. The nearby solution gives a qualitatively different interpretation. At $z=0.028$, the CIGALE fit yields
$M_\star=(3.74\pm1.34)\times10^{7}\,M_\odot$
[$\log_{10}(M_\star/M_\odot)=7.57\pm0.16$] and
${\rm SFR}=0.0017\pm0.0010\,M_\odot\,{\rm yr}^{-1}$, with an SFH age of $7.24\pm2.12$ Gyr. The source would therefore be a low-mass, predominantly old and extremely weakly star-forming system if it is associated with NGC\,4410. In contrast, the intermediate-$z$ solutions imply substantially higher stellar masses and star-formation rates. The inferred physical properties are summarized in Table~\ref{tab:cigale_results}. The nearly two-order-of-magnitude difference in stellar mass between the nearby and intermediate-$z$ interpretations demonstrates that the distance uncertainty is central to determining the nature of the candidate. The CIGALE results should therefore be regarded as redshift-dependent estimates rather than definitive measurements. In the following discussion, we consider the nearby $z=0.028$ and intermediate
$z\simeq0.3-0.35$ interpretations with equal consideration.

\begin{figure*}[t!]
\centering
    \includegraphics[width=0.8\linewidth]{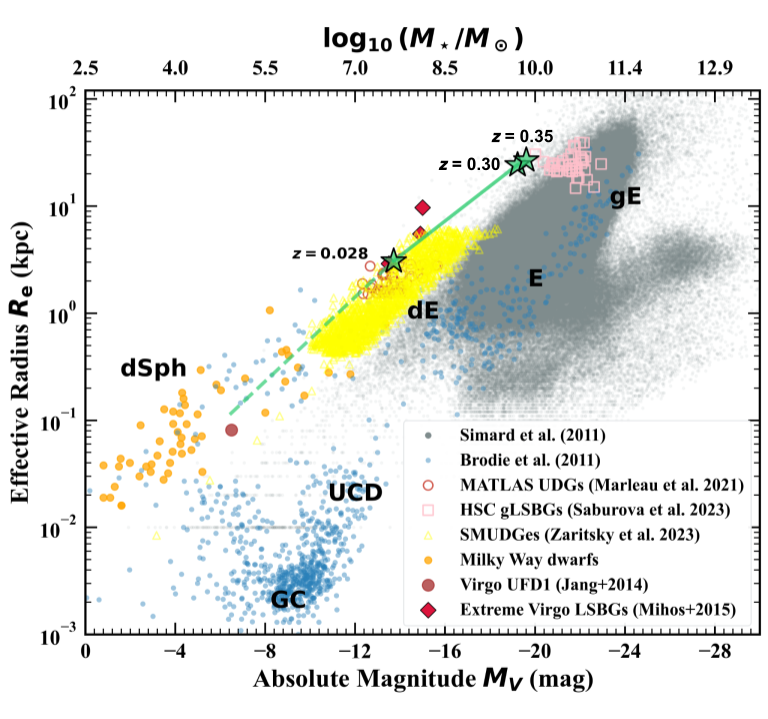}
\caption{Effective radius versus absolute $V$-band magnitude for Rubin J122659.4+090236 under different assumed redshifts. The top axis shows the corresponding stellar mass estimated using the color--$M_{\star}/L$ relation of \citet{2003ApJS..149..289B}. The green track shows the evolution of the inferred physical properties with redshift, with green stars marking the three redshift solutions considered: $z=0.028$, corresponding to the NGC\,4410-associated scenario, and $z=0.30$ and $z=0.35$, corresponding to the unrestricted EAZY solutions without and with the redshift prior, respectively. The background samples show the parameter space occupied by galaxies and stellar systems from \citet{2011ApJS..196...11S}, \citet{2011AJ....142..199B}, MATLAS UDGs \citep{2021AA...654A.105M}, HSC giant LSBGs \citep{2023MNRAS.520L..85S}, SMUDGes UDGs \citep{2023ApJS..267...27Z}, Milky Way dwarf galaxies \citep[][and references therein]{2019ARA&A..57..375S}, Virgo UFD1 \citep{Jang_2014}, and extreme Virgo LSBGs \citep{Mihos_2015}. Labels indicate the approximate loci of globular clusters (GCs), ultra-compact dwarfs (UCDs), dwarf spheroidals (dSphs), dwarf ellipticals (dEs), ellipticals (Es), and giant ellipticals (gEs).}
\label{fig:mv_re}
\end{figure*}

\section{Discussion}
\label{sec:discussion}

The principal uncertainty in the nature of Rubin J122659.4+090236 is its distance. The candidate lies only $2.54\arcmin$ in projection from the interacting NGC\,4410 system at $z\simeq0.02$, while the four-band photometry permits both low-$z$ and intermediate-$z$ solutions. The resulting physical interpretations are illustrated in Figure~\ref{fig:mv_re}, which shows how the candidate moves through the size--luminosity ($M_V$)\footnote{Conversion of LSST photometry to V band magnitude was done using the empirical relation in \url{https://www.sdss3.org/dr8/algorithms/sdssUBVRITransform.php\#Lupton2005}} plane as a function of the assumed redshift. At $z\simeq0.3$--$0.35$, the candidate has $R_e\sim30$~kpc, $M_\star\sim(3$--$4)\times10^9\,M_\odot$, and ${\rm SFR}\sim0.7$--$1.1\,M_\odot\,{\rm yr}^{-1}$. It would therefore be an exceptionally extended background LSBG, with its unusual nature arising primarily from its large physical size and diffuse light distribution. The observed surface-brightness values must, however, be corrected for cosmological dimming and the observed-to-rest-frame bandpass transformation before comparison with local samples (for details see Appendix \ref{app:restframe_sb}). Nevertheless, the source remains intrinsically very low surface brightness. 

Conversely, at $z=0.028$, the candidate has $R_e\simeq3$~kpc and $M_V\sim-14$, placing it among faint, diffuse dwarf-like systems in Figure~\ref{fig:mv_re}. The CIGALE fit gives $M_\star=(3.74\pm1.34)\times10^7\,M_\odot$ and ${\rm SFR}=(1.72\pm1.03)\times10^{-3}\,M_\odot\,{\rm yr}^{-1}$, with an old stellar population. Its red colour, low SFR, shallow Sérsic profile, and proximity to NGC\,4410 are consistent with a diffuse dwarf associated with the group. The source could alternatively be related to the ongoing interaction as a tidal dwarf or intragroup structure \citep{2010AdAst2010E...1B, 2013LNP...861..327D}. A backsplash origin is also possible, since environmental processing can produce red, weakly star-forming systems \citep{2021NatAs...5.1255B}, although the present data cannot establish its orbital history.

Thus, the same observed galaxy admits two fundamentally different physical interpretations: a faint, few-kiloparsec diffuse dwarf potentially associated with NGC\,4410, or an exceptionally extended background LSBG at $z\simeq0.3$--$0.35$. Importantly, its diffuse nature persists in either scenario; the redshift determines the physical scale and origin of that structure. 

\section{Summary and Conclusions}
\label{sec:conclusions}

We report the serendipitous discovery of an exceptionally diffuse LSBG candidate, Rubin J122659.4+090236, in LSST DP2 imaging of the NGC\,4410 field. The source has a smooth, nearly round morphology, a shallow Sérsic profile ($n\simeq0.4$), an effective radius of $\sim6''$, and extremely faint surface brightness, demonstrating the capability of Rubin imaging to reveal diffuse galaxies.

Using the four-band Rubin $ugri$ photometry, we find multiple viable photo-$z$ solutions. The unrestricted EAZY fits yield $z_{\rm peak}=0.30$ without a prior and $z_{\rm peak}=0.35$ with a prior, while a fit restricted to $0<z<0.1$ gives $z_{\rm peak}\simeq0.028$, consistent with NGC\,4410. The sparse photometry therefore does not permit a definitive redshift. At $z\simeq0.028$, the source would be a faint ($M_V\sim-14$), few-kpc diffuse dwarf-like system with $M_\star\simeq3.7\times10^7\,M_\odot$ and very weak star formation, potentially associated with NGC\,4410 and possibly of tidal, intragroup, or backsplash origin. At $z\simeq0.3$--$0.35$, it would instead be an exceptionally extended ($R_e\sim30$ kpc) background LSBG with $M_\star\sim(3$--$4)\times10^9\,M_\odot$ and modest star formation. After accounting for cosmological dimming and bandpass effects, it remains intrinsically diffuse in the intermediate-$z$ scenario. A spectroscopic redshift is needed to distinguish these scenarios, while additional near-infrared photometry could help resolve the redshift degeneracy. This discovery highlights Rubin's potential to uncover extreme LSBGs in the LSST era.

\begin{acknowledgements}
This research uses services or data provided by the Rubin Science Platform at NSF-DOE Vera C. Rubin Observatory, which is jointly funded by the U.S. National Science Foundation and the U.S. Department of Energy, Office of Science. We acknowledge the efforts of the Rubin Observatory staff, the data processing teams, and the broader Rubin collaboration whose work has made these data available. The authors wish to thank the anonymous referee for the valuable comments provided, which helped to improve the manuscript. D.M. acknowledges the help rendered by Dr. Divya Pandey with EAZY code. R.D. gratefully acknowledges support by the ANID BASAL project FB210003. HHT  acknowledges support from project PAPIIT-UNAM IG101725. A.C. has been supported by the Polish National Science Center project UMO-2023/51/D/ST9/00147. Software used in this work includes \textit{Astropy} \citep*{2013astropy,2018astropy,The_Astropy_Collaboration_2022}, \textit{SciPy} \citep{2020SciPy-NMeth}, \textit{NumPy} \citep{van_der_Walt_2011,Harris_2020}, \textit{Matplotlib} \citep{2007CSE.....9...90H}, \texttt{GALFIT} \citep{2002AJ....124..266P}, \textit{EAZY} \citep{2008ApJ...686.1503B}, and \textit{CIGALE} \citep{2019A&A...622A.103B}.
\end{acknowledgements}

\facility{Rubin:Simonyi (LSSTCam), Rubin:USDAC}

\appendix
\restartappendixnumbering
\counterwithin{figure}{section}
\renewcommand{\thefigure}{\thesection\arabic{figure}}

\section{CFHT/NGVS Imaging of Rubin J122659.4+090236}

\label{app:cfht}

\begin{figure*}
\centering
\includegraphics[width=\linewidth]{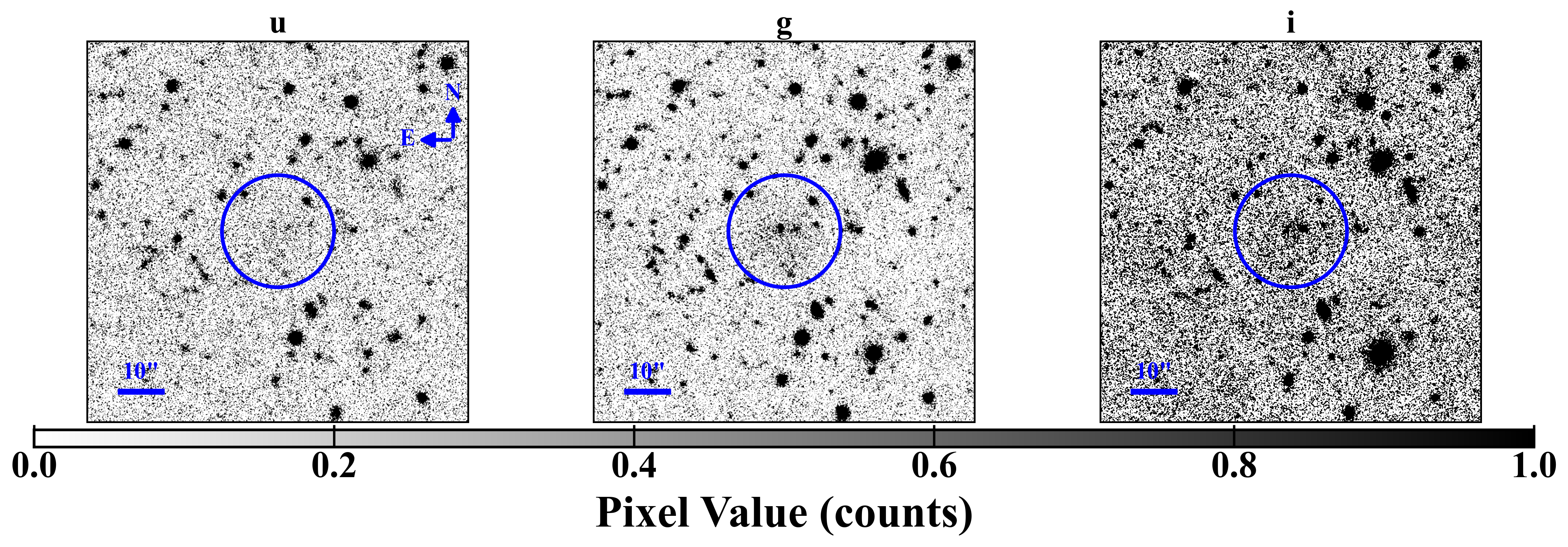}
\caption{Unsharp-masked, using a Gaussian smoothing
scale of $\sigma=8$ pixels ($\simeq1.5\arcsec$), CFHT MegaCam NGVS $u$, $g$, and $i$ cutouts centred on Rubin J122659.4+090236. Each panel covers $81''\times81''$ and uses a common linear intensity scale. The blue circle marks the candidate position, where diffuse emission is detected independently in all three bands, most clearly in $g$. North is up, east is to the left, and the scale bar is $10''$. The mean $3\sigma$ surface-brightness limits measured in $10''\times10''$ apertures are $29.96$, $30.03$, and $28.67$~mag~arcsec$^{-2}$ in $u$, $g$, and $i$, respectively.}
\label{fig:cfht}
\end{figure*}

As an independent check of the Rubin detection, we examined archival Canada--France--Hawaii Telescope (CFHT) MegaCam imaging from the Next Generation Virgo Cluster Survey (NGVS). The candidate is visible as diffuse emission at the same sky position in the available $u$, $g$, and $i$ images (Figure~\ref{fig:cfht}), with the clearest detection in the $g$ band. This independent detection supports the reality of the low-surface-brightness feature seen in the Rubin data. Uniform $r$-band imaging is not available at this position because the NGVS $r$-band observations were incomplete owing to interruptions during the survey \citep{2026arXiv260718414F}. The CFHT cutouts cover $81''\times81''$, matching the Rubin analysis. Using randomly placed $10''\times10''$ background apertures, we estimate mean $3\sigma$ surface-brightness limits of $29.96$, $30.03$, and $28.67$~mag~arcsec$^{-2}$ in the $u$, $g$, and $i$ bands, respectively.

\section{Assessment of background}
\label{app:mask}

As an independent assessment of the background properties of the deep-coadd imaging and the effectiveness of the source masking, we performed a random-aperture analysis directly on the deep-coadd image. Importantly, this test was carried out on the original deep-coadd image after applying the source mask, rather than on the image after subtraction of a locally estimated background using sigma-clipped statistics. Thus, the sky contribution remains present in the aperture measurements, and the resulting flux distribution is not expected to be centred on zero. Instead, the distribution provides an empirical characterisation of the flux fluctuations present in nominally unmasked regions of the image, including the contribution from the residual sky, pixel-scale noise, correlated background fluctuations, and any remaining faint astrophysical emission.

We randomly placed 5000 circular apertures on the unmasked regions of the deep-coadd image, rejecting positions for which the aperture intersected masked pixels and also avoiding the central galaxy region. Two aperture radii were considered, $0.7^{\prime\prime}$ and $1.0^{\prime\prime}$, corresponding to diameters of $1.4^{\prime\prime}$ and $2.0^{\prime\prime}$, respectively. For a spatially well-behaved background, the random-aperture flux distribution is expected to exhibit a relatively well-defined central component, with its location determined primarily by the mean residual sky contribution within the aperture and its width determined by the combined effects of pixel noise and correlated background fluctuations. A positive tail can additionally arise from faint undetected sources, residual low-surface-brightness emission, imperfectly masked source wings, and other positive fluctuations in the image.

\begin{figure*}[t]
    \centering
    \includegraphics[width=0.48\linewidth]{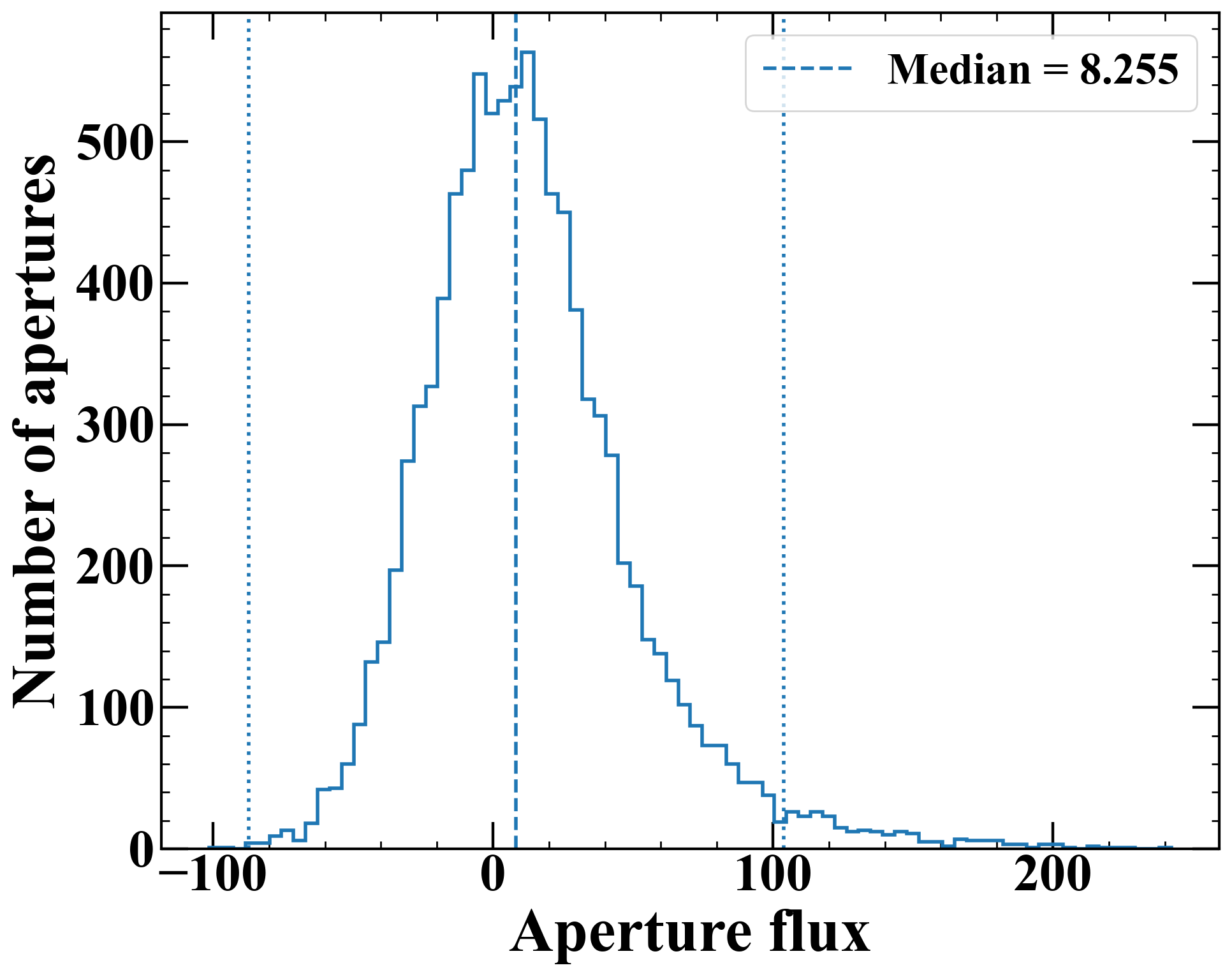}
    \includegraphics[width=0.48\linewidth]{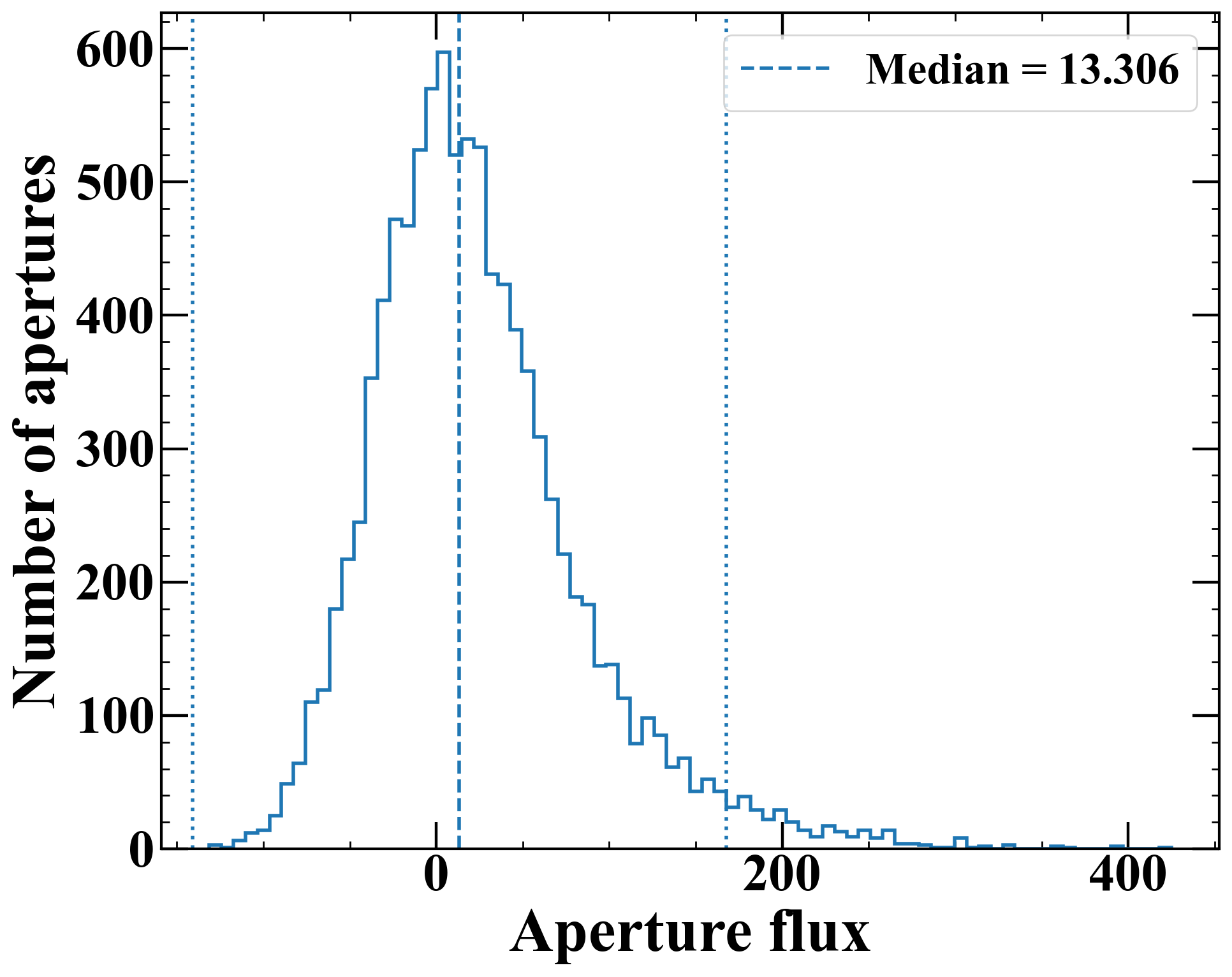}
    \caption{
    Random-aperture flux distributions measured directly from the deep-coadd LSST image after applying the final source mask, without subtracting the sigma-clipped background estimate. In the left panel, 5000 randomly positioned circular apertures with radius $0.7^{\prime\prime}$ (diameter
    $1.4^{\prime\prime}$) are used, while the right panel shows the corresponding distribution for apertures with radius $1.0^{\prime\prime}$ (diameter
    $2.0^{\prime\prime}$). The dashed vertical lines indicate the median
    aperture fluxes of $8.26$~nJy and $13.31$~nJy, respectively, and the dotted lines indicate the corresponding $\pm3\sigma$ ranges. Since the test is performed on the original deep-coadd image, the aperture flux distributions
    retain the contribution from the sky background and are therefore not expected to be centred on zero. The broader distribution for the larger aperture reflects the increased contribution from background fluctuations
    over the larger enclosed area. The extended positive tails likely include contributions from faint undetected sources, residual diffuse emission, and imperfectly masked source wings.
    }
    \label{fig:random_aperture}
\end{figure*}

Figure \ref{fig:random_aperture} shows the distribution of the flux in the apertures. For the $0.7^{\prime\prime}$-radius apertures, the resulting distribution has a well-defined central component with a median flux of $8.26$~nJy, while the $1.0^{\prime\prime}$-radius apertures have a median flux of $13.31$~nJy. The larger aperture also produces a broader flux distribution, as expected because the aperture encompasses a larger number of pixels and therefore integrates over a larger contribution from both random and correlated background fluctuations. The increase in the characteristic aperture flux between the two aperture sizes is likewise expected because a larger aperture contains a greater total contribution from the underlying sky background. The distributions show a relatively concentrated central component together with an extended positive tail. The positive tail is likely to include contributions from faint sources that remain below the detection threshold, residual diffuse emission, and low-level structures not completely removed by the masking procedure. Consequently, the non-zero median aperture fluxes should not be interpreted as evidence for an inadequately subtracted sky background, since the sky has deliberately not been removed for this test. Rather, they quantify the characteristic flux present in nominally blank, unmasked regions of the original deep-coadd image on the two aperture scales considered.

This random-aperture analysis therefore provides a complementary empirical assessment of the background environment of the imaging independently of the particular sigma-clipping procedure adopted to estimate and subtract the background for the subsequent structural analysis. In particular, the relatively well-defined central distributions indicate that the unmasked regions are not dominated by large, spatially varying flux excursions, while the extended positive tails provide an indication of the level of residual contamination from faint or diffuse sources. These measurements are especially relevant for the present analysis because the surface brightness of the outer regions of the target can approach the level of the background fluctuations. The random-aperture distributions consequently provide an empirical estimate of the flux fluctuations encountered on the angular scales relevant to the faintest portions of the surface-brightness measurements.

\section{CIGALE input parameters} 
\label{apcig}
Here, in Table \ref{tabcigale} we present the parameters used for SED fitting using CIGALE.

\begin{table*}[t!]
\caption{Parameter values given as input to CIGALE for SED fitting}
\label{tabcigale}
\begin{tabular}{|ll|}
\hline\hline
\multicolumn{2}{|l|}{\bf{SFH : sfhdelayed - delayed SFH with optional exponential burst}}                                                                                                \\ \hline\hline
\multicolumn{1}{|l|}{e-folding time of the main stellar population (Myr)}    & 500, 800, 1000, 1500, 2000, 3000                                    \\ 
\multicolumn{1}{|l|}{e-folding time of the late starburst population (Myr)}    & 50, 100                                    \\ 
\multicolumn{1}{|l|}{mass fraction of the late starburst population}   & 0.0, 0.001, 0.005                 \\ 
\multicolumn{1}{|l|}{age of the main stellar population (Myr)}               & 1000, 3000, 5000, 7000, 8000, 9000, 10000 \\ 
\multicolumn{1}{|l|}{Normalise the SFH to produce one solar mass}                        & True                                   \\ \hline\hline
\multicolumn{2}{|l|}{\bf{SSP : bc03} \citep{2003MNRAS.344.1000B}}                                                                                                   \\ \hline\hline
\multicolumn{1}{|l|}{Initial mass function (IMF)}                            & Salpeter \citep{1955ApJ...121..161S}                                                 \\ 
\multicolumn{1}{|l|}{Metallicity}                                      & 0.008                                                     \\ \hline\hline
\multicolumn{2}{|l|}{\bf{Dust attenuation :  dustatt\_modified\_CF00} \citep{2000ApJ...539..718C}}                                                                  \\ \hline\hline
\multicolumn{1}{|l|}{V-band attenuation in ISM ($A_{V}^{\rm ISM}$)}                        & 0.001, 0.01, 0.05, 0.1, 0.2, 0.3                                \\ 
\multicolumn{1}{|l|}{$\mu$}                                            & 0.3, 0.44, 0.5                                                      \\ 
\multicolumn{1}{|l|}{power law slope of attenuation in the ISM}        & -0.7                                                     \\ 
\multicolumn{1}{|l|}{power law slope of attenuation in the BCs}         & -0.7, -1, -1.3                                                     \\ \hline\hline
\end{tabular}
\end{table*}

\section{\texttt{GALFIT} Fits with a Simultaneously Fitted Sky}
\label{galskyap}

\begin{figure}
    \centering
    \includegraphics[width=\linewidth]{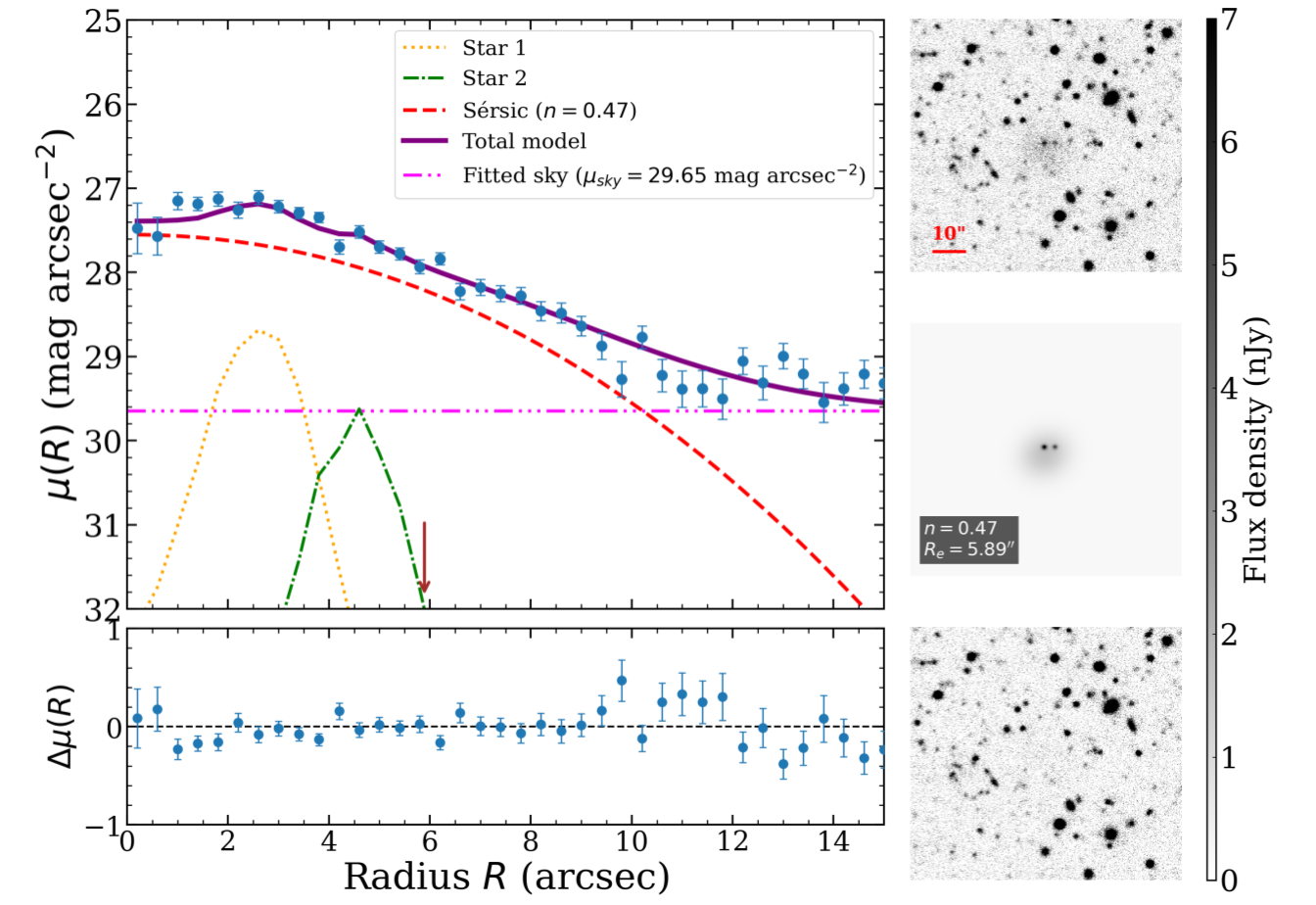}
    \caption{Surface-brightness profile and two-dimensional Sérsic modelling of Rubin J122659.4+090236 in the $g$ band, including the fitted sky background. The upper-left panel shows the observed azimuthally averaged surface-brightness profile (blue points) with the best-fitting Sérsic component (red dashed line), two fitted stellar sources (orange dotted and green dash-dotted lines), and their combined model (purple solid line). The fitted sky level, $\mu_{\rm sky}=29.65$ mag arcsec$^{-2}$, is indicated by the magenta dash-dotted line. The lower-left panel shows the residuals between the observed profile and the total model. The upper-right and lower-right panels show the data and corresponding two-dimensional Sérsic model, respectively. The fit yields $n=0.47$ and $R_e=5.89\arcsec$.}
    \label{figgalfitsky}
\end{figure}

As a check on the adopted background treatment, we repeated the
\texttt{GALFIT} modelling without subtracting the local sky in advance,
allowing a constant sky component to vary simultaneously with the Sérsic
galaxy model and the two PSF components. The resulting structural
parameters are compared with the fiducial sky-subtracted measurements in
Table~\ref{tab:galfit_results}. An example of the resulting $g$-band fit,
including the fitted sky component, is shown in Figure~\ref{figgalfitsky}.

The simultaneous-sky fits give Sérsic indices of
$n=0.47\pm0.03$, $0.46\pm0.02$, and $0.38\pm0.03$ in the $g$, $r$, and
$i$ bands, respectively, compared with $0.45\pm0.03$, $0.43\pm0.02$, and
$0.36\pm0.03$ from the sky-subtracted fits. The effective radii and
surface-brightness parameters are likewise consistent between the two
approaches. The fitted sky surface brightnesses are
$29.65\pm0.02$, $29.11\pm0.02$, and $28.66\pm0.02$
mag~arcsec$^{-2}$ in $g$, $r$, and $i$, respectively. The close agreement
between the two approaches indicates that the shallow Sérsic profiles and
diffuse nature of Rubin J122659.4+090236 are robust to the adopted
treatment of the local sky background.

\section{\texttt{GALFIT} Structural Parameters}
\label{app:galfit}

Table~\ref{tab:galfit_results} provides the complete \texttt{GALFIT} structural parameters for Rubin J122659.4+090236 in the $g$, $r$, and $i$ bands. The table includes the adopted results from the sky-subtracted images together with the corresponding fits in which \texttt{GALFIT} simultaneously models the constant sky background. The two approaches yield consistent structural
parameters, confirming that the inferred diffuse morphology is not strongly dependent on the treatment of the sky background.

\begin{table*}
\centering

\caption{\texttt{GALFIT} structural parameters of Rubin J122659.4+090236 in the
$g$, $r$, and $i$ bands. The adopted measurements are obtained from the
sky-subtracted images and are shown in bold. For comparison, results from
fits in which the sky background was simultaneously fitted by
\texttt{GALFIT} are also listed.}
\label{tab:galfit_results}

\setlength{\tabcolsep}{8pt}
\renewcommand{\arraystretch}{1.0}

\begin{tabular}{lcccccc}
\hline
Parameter &
$g$ (with sky)  & \textbf{$g$} &
$r$ (with sky) & \textbf{$r$ } &
$i$ (with sky) & \textbf{$i$ } \\
\hline

Magnitude (mag) &
$22.07 \pm 0.02$ &
$\mathbf{22.19 \pm 0.02}$ &
$21.48 \pm 0.02$ &
$\mathbf{21.54 \pm 0.02}$ &
$21.30 \pm 0.03$ &
$\mathbf{21.34 \pm 0.03}$ \\

$R_{\rm e}$ (pix) &
$29.46 \pm 1.01$ &
$\mathbf{26.53 \pm 0.90}$ &
$30.00 \pm 0.88$ &
$\mathbf{29.65 \pm 0.85}$ &
$28.48 \pm 1.07$ &
$\mathbf{28.00 \pm 1.04}$ \\

$R_{\rm e}$ (arcsec) &
$5.89 \pm 0.20$ &
$\mathbf{5.31 \pm 0.18}$ &
$6.00 \pm 0.18$ &
$\mathbf{5.93 \pm 0.17}$ &
$5.70 \pm 0.21$ &
$\mathbf{5.60 \pm 0.21}$ \\

$R_{\rm e,circ}$ (pix) &
$28.03 \pm 1.05$ &
$\mathbf{26.38 \pm 0.99}$ &
$28.44 \pm 0.91$ &
$\mathbf{27.71 \pm 0.88}$ &
$26.17 \pm 1.10$ &
$\mathbf{25.73 \pm 1.07}$ \\

$R_{\rm e,circ}$ (arcsec) &
$5.61 \pm 0.21$ &
$\mathbf{5.28 \pm 0.20}$ &
$5.69 \pm 0.18$ &
$\mathbf{5.54 \pm 0.18}$ &
$5.23 \pm 0.22$ &
$\mathbf{5.15 \pm 0.21}$ \\

$n$ &
$0.47 \pm 0.03$ &
$\mathbf{0.45 \pm 0.03}$ &
$0.46 \pm 0.02$ &
$\mathbf{0.43 \pm 0.02}$ &
$0.38 \pm 0.03$ &
$\mathbf{0.36 \pm 0.03}$ \\

$b/a$ &
$0.90 \pm 0.03$ &
$\mathbf{0.99 \pm 0.03}$ &
$0.90 \pm 0.02$ &
$\mathbf{0.87 \pm 0.02}$ &
$0.84 \pm 0.03$ &
$\mathbf{0.84 \pm 0.03}$ \\

$\mu_0$ (mag arcsec$^{-2}$) &
$27.53 \pm 0.10$ &
$\mathbf{27.56 \pm 0.09}$ &
$27.00 \pm 0.08$ &
$\mathbf{27.04 \pm 0.08}$ &
$26.75 \pm 0.11$ &
$\mathbf{26.78 \pm 0.11}$ \\

$\mu_e$ (mag arcsec$^{-2}$) &
$28.20 \pm 0.09$ &
$\mathbf{28.17 \pm 0.08}$ &
$27.63 \pm 0.07$ &
$\mathbf{27.62 \pm 0.07}$ &
$27.21 \pm 0.10$ &
$\mathbf{27.20 \pm 0.10}$ \\

$\langle\mu\rangle_e$ (mag arcsec$^{-2}$) &
$27.81 \pm 0.08$ &
$\mathbf{27.80 \pm 0.08}$ &
$27.25 \pm 0.07$ &
$\mathbf{27.25 \pm 0.07}$ &
$26.89 \pm 0.09$ &
$\mathbf{26.89 \pm 0.09}$ \\

Sky SB (mag arcsec$^{-2}$) &
$29.65 \pm 0.02$ &
-- &
$29.11 \pm 0.02$ &
-- &
$28.66 \pm 0.02$ &
-- \\

\hline
\end{tabular}

\tablecomments{
$R_{\rm e}$ is the effective radius along the semi-major axis, while
$R_{\rm e,circ}=R_{\rm e}\sqrt{b/a}$ is the circularized effective radius.
The Sérsic index is denoted by $n$, and $b/a$ is the axis ratio. The
quantities $\mu_0$, $\mu_e$, and $\langle\mu\rangle_e$ denote the central,
effective, and mean surface brightness within $R_{\rm e}$, respectively.
The sky-fitted columns result from fits in which the sky background was
simultaneously optimized by \texttt{GALFIT}; the corresponding sky surface
brightness is listed in the final row. Bold values denote the adopted
measurements from the sky-subtracted images.
}

\end{table*}
\section{CIGALE Estimated Physical properties}

Table~\ref{tab:cigale_results} summarizes the physical properties derived from the Bayesian CIGALE fits for the three redshift scenarios considered in this work. The table includes the two unrestricted EAZY solutions, with and without the redshift prior, and the restricted $z=0.028$ solution
corresponding to the possible NGC\,4410 association. The inferred stellar mass and star-formation rate vary substantially between the nearby and intermediate-$z$ interpretations, highlighting the strong dependence
of the derived physical properties on the uncertain distance. Uncertainties listed in the table are the Bayesian uncertainties reported by CIGALE.

\begin{table*}
\centering
\setlength{\tabcolsep}{20pt}
\caption{Physical properties of Rubin J122659.4+090236 derived from the Bayesian CIGALE fits for the three photo-$z$ scenarios considered in this work. The first two columns correspond to the unrestricted EAZY fits without and with the redshift prior, respectively, while the final column represents the restricted $0<z<0.1$ solution testing association with NGC\,4410.}
\label{tab:cigale_results}
\begin{tabular}{lccc}
\hline
Parameter & $z=0.30$ & $z=0.35$ & $z=0.028$ \\
 & No prior & With prior & NGC\,4410 scenario \\
\hline
$D_L$ (Mpc)
& 1577.9 & 1834.4 & 121.9 \\
SFH age (Gyr)
& $4.47\pm2.01$ & $3.88\pm1.86$ & $7.24\pm2.12$ \\
$M_\star$ ($M_\odot$)
& $(2.98\pm1.28)\times10^{9}$
& $(3.48\pm1.57)\times10^{9}$
& $(3.74\pm1.34)\times10^{7}$ \\
$\log_{10}(M_\star/M_\odot)$
& $9.47\pm0.19$
& $9.54\pm0.20$
& $7.57\pm0.16$ \\
$M_{\star,\rm old}$ ($M_\odot$)
& $(2.97\pm1.28)\times10^{9}$
& $(3.47\pm1.57)\times10^{9}$
& $(3.74\pm1.34)\times10^{7}$ \\
$M_{\star,\rm young}$ ($M_\odot$)
& $(7.20\pm2.43)\times10^{6}$
& $(1.11\pm0.40)\times10^{7}$
& $(1.68\pm1.00)\times10^{4}$ \\
SFR ($M_\odot\,{\rm yr}^{-1}$)
& $0.74\pm0.25$
& $1.14\pm0.41$
& $0.0017\pm0.0010$ \\
$M_{\rm gas}$ ($M_\odot$)
& $(1.10\pm0.57)\times10^{9}$
& $(1.25\pm0.68)\times10^{9}$
& $(1.57\pm0.63)\times10^{7}$ \\
$A_V^{\rm BC}$ (mag)
& $0.158\pm0.182$
& $0.170\pm0.191$
& $0.164\pm0.185$ \\
$A_V^{\rm ISM}$ (mag)
& $0.103\pm0.105$
& $0.109\pm0.108$
& $0.107\pm0.107$ \\
\hline
\end{tabular}
\end{table*}

\section{Rest-frame Surface Brightness at Intermediate Redshift}
\label{app:restframe_sb}

The interpretation of the observed surface brightness changes with the adopted redshift because of cosmological surface-brightness dimming and the mapping between the observed and rest-frame bandpasses. We consider both intermediate-$z$ solutions obtained from EAZY, namely $z=0.30$ without a redshift prior and $z=0.35$ with the EAZY prior.

For a surface brightness expressed as a specific intensity per unit frequency, the cosmological dimming scales as $(1+z)^{-3}$ \citep{2001AJ....122.1084L}. The corresponding AB surface-brightness correction is therefore
\begin{equation}
\mu_{\rm rest} =
\mu_{\rm obs} - 7.5\log_{10}(1+z).
\end{equation}
This gives corrections of $0.86$ and $0.98$ mag for $z=0.30$ and
$z=0.35$, respectively. The resulting surface-brightness values are listed in Table~\ref{tab:restframe_sb}.

At $z=0.30$, the Rubin $r$-band pivot wavelength
($\lambda_{\rm piv}=6223$~\AA) corresponds to
$6223/(1+z)=4772$~\AA, remarkably close to the rest-frame $g$-band pivot of $4827$~\AA. The observed $r$ band therefore provides an almost model-independent estimate of the rest-frame $g$-band surface brightness, with only a small differential $k$-correction. From Table~\ref{tab:restframe_sb}, the corrected values are $\mu_{0,g}^{\rm rest}=26.18$,
$\mu_{e,g}^{\rm rest}=26.76$, and
$\langle\mu_e\rangle_g^{\rm rest}=26.39$
mag~arcsec$^{-2}$. Thus, even after correcting for cosmological dimming, the candidate remains an intrinsically very low-surface-brightness system.

For the prior-informed solution, $z=0.35$, the observed $r$ band samples $\simeq4623$~\AA\ in the rest frame and is therefore less closely matched to rest-frame $g$. The corrected $r$-band values in Table~\ref{tab:restframe_sb} should consequently be regarded as approximate rest-frame optical surface brightnesses, before applying any differential $k$-correction.

The observed $g$ band provides a less direct comparison at these
redshifts because it samples rest-frame wavelengths blueward of the $4000$~\AA\ break. Applying only the cosmological correction to the observed $g$ band therefore does not yield an exact rest-frame measurement. The close $r\rightarrow g$ correspondence at $z=0.30$ provides the cleanest and least model-dependent comparison with local $g$-band LSBG measurements.

The observed colours, $g-r=0.55$ and $g-i=0.76$, are likewise not
rest-frame colours at either intermediate-$z$ solution. At
$z\simeq0.3$--$0.35$, the Rubin bands sample rest-frame wavelengths that straddle the $4000$~\AA/Balmer-break region, so quantitative rest-frame colours require a $k$-correction based on the galaxy SED. Under the nearby $z\simeq0.028$ interpretation, in contrast, the observed bands are much closer to the rest frame and the colours can be compared more
directly with local LSBG populations.

\begin{table*}
\centering
\caption{Observed and cosmologically corrected surface-brightness
measurements for Rubin J122659.4+090236 under the two intermediate-$z$
solutions. The correction assumes surface brightness expressed as specific
intensity per unit frequency, for which the cosmological dimming scales as
$(1+z)^3$. No differential $k$-correction is applied. At $z=0.30$, the
observed $r$ band closely samples rest-frame $g$ and therefore provides the
most direct comparison with local $g$-band measurements.}
\label{tab:restframe_sb}

\setlength{\tabcolsep}{8pt}
\renewcommand{\arraystretch}{1.15}

\begin{tabular}{lccccccccc}
\hline
& \multicolumn{3}{c}{$g$ band} &
  \multicolumn{3}{c}{$r$ band} &
  \multicolumn{3}{c}{$i$ band} \\
\cline{2-10}
Quantity &
Observed & $z=0.30$ & $z=0.35$ &
Observed & $z=0.30$ & $z=0.35$ &
Observed & $z=0.30$ & $z=0.35$ \\
\hline

$\mu_0$ &
27.56 & 26.70 & 26.58 &
27.04 & 26.18 & 26.06 &
26.78 & 25.92 & 25.80 \\

$\mu_e$ &
28.17 & 27.31 & 27.19 &
27.62 & 26.76 & 26.64 &
27.20 & 26.34 & 26.22 \\

$\langle\mu\rangle_e$ &
27.80 & 26.94 & 26.82 &
27.25 & 26.39 & 26.27 &
26.89 & 26.03 & 25.91 \\

\hline
\end{tabular}

\tablecomments{
The cosmological corrections are $0.86$ mag at $z=0.30$ and
$0.98$ mag at $z=0.35$, corresponding to
$7.5\log_{10}(1+z)$. These corrections account only for cosmological
surface-brightness dimming; the corrected values should therefore not be
interpreted as exact rest-frame surface brightnesses when the observed and
rest-frame filters are not directly matched. At $z=0.30$, the observed
$r$-band pivot wavelength of $6223$~\AA\ corresponds to $4772$~\AA\ in the
rest frame, close to the rest-frame $g$ pivot of $4827$~\AA. The corrected
$r$-band values therefore provide the most model-independent estimate of
the rest-frame $g$-band surface brightness.
}
\end{table*}

\bibliography{ref}{}
\bibliographystyle{aasjournalv7}



\end{document}